\documentclass[aps,prd,groupedaddress,amssymb,twocolumn,eqsecnum,showpacs,epsfig,nofootinbib]{revtex4}

\usepackage{hyperref}
\usepackage{graphicx}
\usepackage{bm}
\usepackage{dcolumn}
  
\usepackage{amsmath}
\usepackage{amssymb} 

\numberwithin{equation}{section}
\usepackage{epsfig}

\begin{document}
\newcommand{\newc}{\newcommand}

\newc{\be}{\begin{equation}}
\newc{\ee}{\end{equation}}
\newc{\bear}{\begin{eqnarray}}
\newc{\eear}{\end{eqnarray}}
\newc{\bea}{\begin{eqnarray*}}
\newc{\eea}{\end{eqnarray*}}
\newc{\D}{\partial}
\newc{\ie}{{\it i.e.} }
\newc{\eg}{{\it e.g.} }
\newc{\etc}{{\it etc.} }
{\newc{\etal}{{\it et al.}}
\newc{\lcdm}{$\Lambda$CDM}
\newcommand{\nn}{\nonumber}
\newc{\ra}{\rightarrow}
\newc{\lra}{\leftrightarrow}
\newc{\lsim}{\buildrel{<}\over{\sim}}
\newc{\gsim}{\buildrel{>}\over{\sim}}
\newcommand{\mincir}{\raise
-3.truept\hbox{\rlap{\hbox{$\sim$}}\raise4.truept\hbox{$<$}\ }}
\newcommand{\magcir}{\raise
-3.truept\hbox{\rlap{\hbox{$\sim$}}\raise4.truept\hbox{$>$}\ }}

\title{Growth index of matter perturbations in the light of Dark Energy Survey}



\author{Spyros Basilakos}\email{svasil@academyofathens.gr}
\affiliation{Academy of Athens, Research Center for Astronomy and
Applied Mathematics,
 Soranou Efesiou 4, 11527, Athens, Greece}
\affiliation{National Observatory of Athens, V. Paulou and I. Metaxa 15236, Penteli, Greece}

\author{Fotios K. Anagnostopoulos} \email{fotis-anagnostopoulos@hotmail.com}
\affiliation{{Department of Physics, National \& Kapodistrian University of Athens, Zografou Campus GR 157 73, Athens, Greece}}

\begin{abstract}
We study how the cosmological constraints from growth data 
are improved by including the measurements of bias from Dark 
Energy Survey (DES). In particular, we utilize the 
biasing properties of the DES
Luminous Red Galaxies (LRGs)
and the growth data provided by the various galaxy surveys
in order to constrain the growth index ($\gamma$) of the linear
matter perturbations.
Considering a
constant growth index we can put tight constraints, up to
$\sim 10\%$ accuracy, on $\gamma$. 
Specifically, using the priors of the Dark Energy Survey
and implementing a joint likelihood procedure 
between theoretical expectations and data we find
that the best fit value is in between 
$\gamma=0.64\pm 0.075$
and $0.65\pm 0.063$. On the other hand utilizing the Planck priors 
we obtain $\gamma=0.680\pm 0.089$
and $0.690\pm 0.071$. 
This shows a small but non-zero deviation from 
General Relativity ($\gamma_{\rm GR}\approx 6/11$), 
nevertheless 
the confidence level is in the range $\sim 1.3-2\sigma$.
Moreover, we find that the estimated mass of the 
dark-matter halo in which LRGs survive 
lies in the interval 
$\sim 6.2 \times 10^{12} h^{-1} M_{\odot}$ and $1.2 \times
10^{13} h^{-1} M_{\odot}$, for the different bias models. 
Finally, allowing $\gamma$ to evolve with redshift 
[Taylor expansion: $\gamma(z)=\gamma_{0}+\gamma_{1}z/(1+z)$] we find that 
the $(\gamma_{0},\gamma_{1})$ parameter solution space 
accommodates the GR prediction 
at $\sim 1.7-2.9\sigma$ levels.

\end{abstract}
\maketitle

\section{Introduction}
The past and present analysis of various cosmological data
(SNIa, Cosmic Microwave Background-CMB, Baryonic Acoustic
Oscillations-BAOs, Hubble parameter measurements etc) 
converge to the following cosmological paradigm,
the observed Universe is spatially flat and the cosmic fluid 
consists of $\sim 4\%$ luminous (baryonic) matter, $\sim 26\%$
dark matter and $\sim 70\%$ some sort of dark energy (hereafter DE) which 
plays a key role in explaining the accelerated expansion of the
universe (cf. \cite{Hicken2009,Komatsu2011,Blake2011,Hinshaw2013,Farooq2013,Ade2013,Aghanim:2018eyx} and references therein).
Despite the fact that there is an agreement among the majority of
cosmologists concerning the
ingredients of the cosmic fluid however, there are different explanations 
regarding the physical mechanism which causes 
the accelerated expansion of the
universe. In brief, the general avenue that one can design
in order to study cosmic acceleration 
is to treat DE either as a new field in nature or as a
modification of General Relativity (see for review
\cite{Copeland2006,Caldwell2009,Amendola2010}).

It has been proposed 
\cite{Linder2004,Linder2007,Mar2014,Bas2016} that
in order to discriminate between scalar field
DE and modified gravity one may utilize the evolution of the
linear growth of matter fluctuations
$\delta_{m} (z)=\delta \rho_m / \rho_m$.
In particular, we introduce the growth rate of clustering,
which is given by 
$f(a)=\frac{d{\rm ln}D}{d{\rm ln}a}\simeq \Omega_{m}^{\gamma}(a)$,
where $D(a)=\delta_{m}(a)/\delta_{m}(a=1)$
is the linear growth factor (normalized to unity at the present epoch),
$a(z)=(1+z)^{-1}$ is the scale factor of the universe,
$\Omega_m(a)$ is the dimensionless matter density parameter and $\gamma$
is the so called growth index \cite{Peebles1993,Wang1998}.
In fact the determination of the growth index is considered one of
the main targets in these kind of studies because 
it can be used in order to test 
General Relativity (GR) on extragalactic scales, even in a model independent
fashion \cite{Ness2015}.
Indeed,  in  the  literature  one  may  find
a  large  family  of  studies  in  which  the  functional form
of the growth index is given analytically for several
cosmological  models namely, scalar  field  DE  
\cite{Linder2007,Wang1998,Silveira1994,Nesseris2008,Basilakos2012}, 
DGP \cite{Linder2007},\cite{Wei2008,Gong2008,Fu2009},
$f(R)$ \cite{Gannouji2009,Tsujikawa2009}, $f(T)$ \cite{BasFT}
Finsler-Randers \cite{Basilakos2013}, running vacuum models
\cite{Basola2015}, clustered and Holographic dark energy \cite{Mehra2015}. 

From the view point of large scale structure, the study of
the distribution of matter on extragalactic 
using different mass tracers
(galaxies, AGNs, clusters
of galaxies etc) provides important constraints on theories of 
structure formation. Specifically,
owing to the fact that gravity reflects, via gravitational instability,
on the physics of clustering \cite{Peebles1993}
it is natural to utilize
the clustering/biasing properties
of the extragalactic mass tracers in constraining cosmological models
(see \cite{Matsubara2004,Basilakos2005,Basilakos2006,
Krumpe2013})
as well as to test
the validity of GR on cosmological scales
(see \cite{Basilakosetal2012},\cite{Bean2013}).
Following the above lines, in the current article we 
combine the linear bias data of Luminous Red Galaxies
(hereafter LRGs; \cite{DES2017}), recently released by the DES group, with 
the growth rate data as provided by Sargedo {\it et al}. \cite{Sarg},
in order to place constraints on $(\gamma,M_{h})$. Notice that 
$M_{h}$ is the dark matter halo in which the LRG live.

The structure of the paper is as follows.
In section II we present the DES bias data and the growth data.
In section III we provide the family of basic bias models, while in 
section IV we discuss the evolution of linear matter fluctuations.
The outcome of our analysis 
is presented in section V, while our main conclusions can be found 
in section VI.

\section{DESY1 Red Galaxies Bias Data and Growth data}
In a sequence of previous theoretical articles 
we have proposed to use the biasing properties
of extragalactic sources in order to 
constrain the growth index of matter 
fluctuations \cite{Basilakosetal2012}. 
Therefore, in the light of 
recent Dark Energy Survey (DES) bias data, we attempt to compare 
the predictions of the most popular linear bias models (see below) 
with the data. Specifically, the 
DES bias data \cite{DES2017} were 
extracted in the context of 
the angular correlation function (ACF)
using the 1-year DES sample 
of $\sim 6.6\times10^{5}$ LRGs as tracers of the LSS. 
This population of galaxies can be observed 
$B$ corresponds to DES, 
In Table \ref{tab01} we list the numerical
values of the DES bias data with the corresponding errors. 

The aim of our work is the following:
if we accept that the background expansion is given by the concordance 
$\Lambda$CDM model then we are interested to 
check the growth index of matter fluctuations.
Specifically, 
we restrict the present analysis to the most popular expansion models. First 
we utilize the DES/Planck/JLA/BAO $\Lambda$CDM cosmology, namely
$\Omega_{m0}=1-\Omega_{\Lambda 0}=0.301$, $h=0.682$, $\Omega_{b0}=0.048$, $n=0.973$ 
with $\sigma_{8}=0.801$ \cite{Abbot2017} and second we use the 
Planck TT+TE+EE+low+lensing $\Lambda$CDM cosmology, hence
$\Omega_{m0}=1-\Omega_{\Lambda0}=0.3153$, $h=0.6736$, 
$\Omega_{b0}=0.0493$, $n=0.9649$, and $\sigma_{8}=0.811$
\cite{Aghanim:2018eyx}. 
In this context, the normalized Hubble parameter of the $\Lambda$CDM model
is written as 
\begin{equation}
\label{Eaa} 
E(z)=\left[\Omega_{m0}(1+z)^{3}+\Omega_{\Lambda 0}\right]^{1/2} \;.
\end{equation}



\begin {table}
\begin{center}
    \begin{tabular}{ c c  c  }
    \hline 
    \\[0.1ex]
Red. Range&    Median Redshift     & DESY1 bias       \\[1.ex]  \hline 
$0.15<z<0.3$   & $0.225\pm 0.075$ &  $1.40\pm0.077$   \\     
$0.3<z<0.45$   & $0.375\pm 0.075$ &  $1.61\pm0.051$   \\
$0.45<z<0.6$   & $0.525\pm 0.075$ &  $1.60\pm0.040$   \\ 
$0.6<z<0.75$   & $0.675\pm 0.075$ &  $1.93\pm0.045$   \\ 
$0.75<z<0.9$   & $0.825\pm 0.075$ &  $1.99\pm0.066$   \\  \hline 
    \end{tabular}
\end{center}
\caption {The measured bias data 
of the 1-year DES LRGs from Elvin-Poole et. al. \cite{DES2017}.}
\label{tab01}
\end{table}

In addition to DES bias data, we use 
in our analysis the growth data and the corresponding covariances
as collected by 
Sargedo {\it et al}. \cite{Sarg} (see their Table I and references therein).
This sample contains 22 entries 
for which the product $f(z)\sigma_{8}(z)$
is available as a function of redshift, where 
$f(z)$ is the growth rate of clustering\footnote{By definition the 
estimator $f(z)\sigma_{8}(z)$ 
is independent from linear bias and thus it is not 
affected by the dark matter halo. Indeed,
the observed growth rate of structure ($f_{\rm obs}(z)=\beta b$) is
derived from the redshift space distortion parameter
$\beta(z)$ and the linear bias. Observationally,
using the anisotropy of the spatial correlation function one can estimate the
$\beta(z)$ parameter. The linear bias 
factor can be defined as the ratio of
the variances of the tracer (galaxies, Luminous Red  Galaxies etc) 
and underlying mass density fields, smoothed
at  $8h^{-1}$Mpc $b(z)  =\sigma_{8,\rm tr}(z)/\sigma_{8}(z)$,  where
$\sigma_{8,\rm tr}(z)$ is  measured  directly  from  the  sample.
Combining  the  above  definitions we  arrive  at
$f\sigma_{8}=\beta \sigma_{8,\rm tr}$, hence 
the growth rate data are not not affected from bias, 
implying that the two data sets used in the present analysis are not correlated.
}. It is well known that the product
$f\sigma_{8}$ is almost a model-independent
parametrization of expressing the observed growth history
of the universe \cite{Song09}. 



\section{Bias Models}
Let us first briefly present the main bias models. 
In particular, from the so called 
merging bias family we include here the models of
Sheth, Mo \& Tormen \cite{She2001}, 
Jing \cite{Jing1998}
and De Simeone {\it et al}. \cite{deSim2011}. 

For these models 
the bias factor is written as a function of the peak-height parameter, $\nu =
\delta_{c}(z)/\sigma(M_h,z)$ where $\delta_{c}$ is the linearly
extrapolated density threshold above which structures collapse. 
Here we use the accurate fitting formula 
of Weinberg \& Kamionkowski \cite{Wein} to estimate $\delta_{c}(z)$.
Moreover, the mass variance is written as

\begin{equation}\label{eq:wp18}
\sigma(M_h,z)=\left[\frac{D^2(z)}{2\pi^2}\int_0^{\infty}k^2P(k)W^2(kR)dk\right]^{1/2}
\end{equation}
where $W(kR) = 3[sin(kR) - kRcos(kR)]/(kR)^3$ is 
the top-hat smoothing kernel with 
$R = [3M_h/(4\pi\rho_{m})]^{1/3}$, $M_h$ is the halo mass 
and $\rho_{m}$ is the present value of the 
mean matter density, namely 
$\rho_{m}\simeq 2.78 \times 10^{11}\Omega_{m}M_{\odot}$/Mpc$^{3}$.
The quantity $P(k, z)$ is the CDM linear power 
spectrum given by 
$P(k)=P_0k^nT^2(k)$ where 
$n$ is the spectral index of the primordial power 
spectrum and $T(k)$ is the CDM transfer function 
provided by \cite{EisensteinHu}: 

\begin{equation}\label{eq:wp16}
T(k) = \frac{L_0}{L_0 +C_0q^2}
\end{equation}
with $L_0={\rm ln}(2e+1.8q)$, $e=2.718$, $C_0 = 14.2+\frac{731}{1+62.5q}$ 
and $q =k/\Gamma$ with $\Gamma$ being 
is the shape parameter given by \cite{Sugiyama1995}:

\begin{equation}
\Gamma= \Omega_{m}h{\rm exp}(-\Omega_{b}-\sqrt{2h}\frac{\Omega_{b}}{\Omega_{m}}).
\end{equation}
Taking the aforementioned quantities into account and using 
Eq.(\ref{eq:wp18}) the normalization of the power spectrum becomes
\begin{equation}\label{eq:wp19}
P_0 = 2\pi^2\sigma_8^2\left[\int_0^{\infty}T^2(k)k^{n+2}W^2(kR_8)dk\right]^{-1}
\end{equation}
where $\sigma_{8}\equiv \sigma(R_{8},0)$.

Below we provide some details concerning the bias models.

\subsection{SMT and JING}
Sheth, Mo \& Tormen (\cite{She2001}, hereafter SMT) based on 
the ellipsoidal collapse model they found the following bias 
formula 

\begin{equation}\label{eq:wp26}
b(\nu) = 1 + \frac{1}{\sqrt{\alpha}}\delta_c(z)[\sqrt{\alpha}(\alpha\nu^2)
+ \sqrt{\alpha}b(\alpha\nu^2)^{1-c} - f(\nu)]
\end{equation}
with
\begin{equation}\label{eq:wp27}
f(\nu) = \frac{(\alpha\nu^2)^c}{(\alpha\nu^2)^c + b(1-c)(1 - c/2)}\;.
\end{equation}
Using N-body simulations they evaluated 
the free parameters of the model, $\alpha = 0.707, b = 0.5, c = 0.6$

Also, Jing \cite{Jing1998} proposed the following 
bias form

\begin{equation}\label{eq:wp101}
b(\nu) = \left(\frac{0.5}{\nu^4}+1\right)^{0.06-0.02\nu}\left(1+\frac{\nu^2-1}{\delta_c}\right).
\end{equation}



\subsection{DMR}



De Simone et. al. \cite{deSim2011} (hereafter DMR) 
generalized the original Press-Schether 
formalism incorporating a non-Markovian extension with a
stochastic barrier. In this model, 
the critical value for spherical collapse was assumed to be 
a stochastic variable, 
whose scatter reflects a number of complicated aspects of the
underlying dynamics. 
Therefore, the bias factor is 

\begin{multline}\label{eq:wp200}
b(\nu) = 1 + \sqrt{\alpha}\frac{\nu^2}{\delta_c}\left[1+0.4\left(\frac{1}{\alpha\nu^2}\right)^{0.6}\right] \\
-\frac{1}{\sqrt{\alpha}\delta_c\left[1+0.067\left(\frac{1}{\alpha\nu^2}\right)^{0.6}\right] }\;.
\end{multline}

\subsection{BPR}
In addition to merging bias models we shall use the 
generalized model of Basilakos {\it et al}.
\cite{Basilakosetal2012} (hereafter BPR). This 
form of bias is 
valid for any dark energy model 
including those of modified gravity.
In this case, using the hydrodynamic equations of motion, linear 
perturbation theory 
and the Friedmann-Lemaitre solutions a 
second differential equation of bias is derived 
\cite{Basilakosetal2012}.
The solution of the differential equation is given by:
\begin{equation}\label{eq:wp23}
 b(z) = 1 + \frac{b_0 - 1}{D(z)} + C_2 \frac{J(z)}{D(z)}
\end{equation}
with $ J(z) = \int_{0}^{z} \frac{1+y}{E(y)}dy $ 
where $b_0$ is the bias factor at the present time. 
The integration constants $b_0$ and $C_2$ can be found in 
\cite{Basilakosetal2012}, namely

\begin{equation}\label{eq:wp24}
b_0 = 0.857\left[1+\left(\frac{\Omega_{m}}{0.27}\frac{M_h}{10^{14}h^{-1}M_{\odot}}\right)^{0.55}\right]
\end{equation}
and
\begin{equation}\label{eq:wp25}
C_2 = 1.105\left(\frac{\Omega_{m}}{0.27}\frac{M_h}{10^{14}h^{-1}M_{\odot}}\right)^{0.255}\;.
\end{equation}

\section{Evolution of liner growth}
In this section we discuss the main points of the linear growth 
of matter fluctuations
via which the growth index,
$\gamma$, enters in the current analysis.
Focusing on sub-horizon scales 
the differential equation that governs the linear matter perturbations 
(\cite{Linder2004,Linder2007,Lue2004,Stabenau2006,Uzan2007,Tsujikawa2008} 
and references therein) is given by
\be
\label{odedelta}
\ddot{\delta}_{m}+ 2H\dot{\delta}_{m}=4 \pi G_{\rm eff} \rho_{m} \delta_{m},
\ee
where $\rho_{m}\propto a^{-3}$ is the matter density,
$G_{\rm eff}= G_N Q(t)$ with $G_N$ being the
Newton's gravitational constant, while the effects of modified
gravity are encapsulated in the quantity $Q(t)$. 
Of course for those DE models which adhere to General Relativity 
$G_{\rm eff}$ reduces to $G_{N}$, hence $Q(a)=1$.

The solution of the aforementioned equation (\ref{odedelta}) is
$\delta_{m}\propto D(a)$, where $D(a)$ is the growth factor.
For any type of gravity the growth rate of clustering 
is given by the following useful parametrization \cite{Peebles1993,
Wang1998,Linder2007}
\be \label{fa}
 f(a)=\frac{d{\rm ln} \delta_m}{d{\rm ln}a} \simeq \Omega_{m}^{\gamma}(a)
\ee
and thus we have 
\begin{equation}
\label{Dz221}
D(a)={\rm exp} \left[\int_{1}^{a(z)} \frac{\Omega_{m}^{\gamma}(y)}{y}dy \right],
\end{equation}
where $\Omega_{m}(a)=\Omega_{m0}a^{-3}/E^{2}(a)$ and $\gamma$ is the growth index.
Notice that the growth factor is
normalized to unity at the present epoch.

Now, inserting the operator $d/dt=H \; d/d\ln a$ and Eq.(\ref{fa}) 
into Eq.(\ref{odedelta}) we arrive at
\be
\label{faaa}
\frac{df}{d{\rm ln}a}+ f^2 +\left(\frac{\dot H}{H^2}+2\right)f=
\frac{3}{2} Q(a)  \Omega_{m}(a).
\ee
Considering the concordance $\Lambda$CDM model, namely 
$Q(a)=1$ it is easy to show that 
\be
\frac{\dot H}{H^{2}}+2=\frac{1}{2}-\frac{3}{2}{\rm
  w}(a)\left[1-\Omega_{m}(a)\right],
\ee
where $w(a)=-1$.
In this case the Hubble parameter $H(a)=H_{0}E(a)$, where $E(a)$ 
is given by Eq.(\ref{Eaa}) 
and $H_{0}$ is the Hubble
constant\footnote{For the comoving distance
  and for the dark matter halo mass we use the traditional
  parametrization $H_{0}=100h$km/s/Mpc. Of course, when we treat the
  power spectrum shape parameter $\Gamma$ we utilize
  $h\equiv {\tilde h}=0.68$ \cite{Abbot2017}.}.

Generally speaking the growth index
may not be a constant but rather evolve with
redshift; $\gamma\equiv \gamma(z)$. In this framework, 
substituting 
Eq.(\ref{fa}) into Eq.(\ref{faaa}) we find
\begin{equation}
\label{agamz}
-(1+z)\gamma^{\prime}{\rm ln}(\Omega_{m})+\Omega_{m}^{\gamma}+ 3{\rm w}   
(1-\Omega_{m})\left(\gamma-\frac{1}{2}\right)+\frac{1}{2}=\frac{3}{2}Q\Omega_{m}^{1-\gamma}
\end{equation}
where the prime denotes derivative with respect to redshift.
In the present work we restrict our analysis to the following 
two cases \cite{Basilakos2012,Polarski2008,Bal08,Belloso2009}:
\begin{equation}
\gamma(z)=\left\{ \begin{array}{cc}
       \gamma_{0}, &
       \mbox{$\Gamma_{1}$-parametrization}\\
       \gamma_{0}+\gamma_{1}z/(1+z),& \mbox{$\Gamma_{2}$-parametrization.}
       \end{array}
        \right.
\end{equation}
Using the latter $\Gamma_{2}$-parametrization, 
which is nothing else but a Taylor expansion around $a(z)=1$, together
with Eq.(\ref{agamz}) evaluated at the
present time ($z=0$), we can write 
the parameter $\gamma_{1}$ in terms of $\gamma_{0}$
\begin{equation}
\label{Poll2}
\gamma_{1}=\frac{\Omega_{m0}^{\gamma_{0}}+3{\rm w}_{0}(\gamma_{0}-\frac{1}{2})
(1-\Omega_{m0})-\frac{3}{2}Q_{0}\Omega_{m0}^{1-\gamma_{0}}+\frac{1}{2}  }
{\ln  \Omega_{m0}}\;.
\end{equation}
At large redshifts ($z\gg 1$) $\Omega_{\Lambda}(z)\simeq 0$ 
the asymptotic value of the
growth index becomes $\gamma_{\infty}=\gamma_{0}+\gamma_{1}$. 
In general, plugging $
\gamma_{0}=\gamma_{\infty}-\gamma_{1}$ 
into Eq. (\ref{Poll2}) we can define the constants $\gamma_{1}$
as a function of $(\Omega_{m0},\gamma_{\infty},w_{0},Q_{0})$.
For examble, in the case of $\Omega_{m0}=0.301$, $\gamma_{\infty}\simeq 6/11$, 
$w_{0}=-1$ and $Q_{0}=1$, the above calculations give 
$\gamma_{0}^{(th)}\simeq 0.556$.
$\gamma_{1}^{(th)}\simeq -0.011$.

\section{The Likelihood analysis}
In this section we provide the
statistical method 
that we adopt in order to constrain the growth index,
presented in the previous section. 
We implement a standard $\chi^{2}$ minimization
analysis in order to constrains either the
$(\gamma,M_{h})$ parameter space.
 Specifically, in our case the situation is as follows:

\noindent
{\bf (1)} For the DES biasing cosmological probe we use
\be
\label{eq:likel}
\chi^{2}_{\rm DES}({\bf p}_{1})=
\sum_{i=1}^{5} \left[ \frac{b_{\rm obs}(z_{i})-
b_{th}(z_{i},{\bf p}_{1})}
{\sigma_{bi}}\right]^{2}
\ee
where the various forms of the theoretical $b_{\rm th}$ 
are given in section III. Notice that 
$\sigma_{bi}=\sqrt{\sigma_{i}^{2}+\sigma_{z}^{2}}$, where 
$\sigma_{i}$ and $\sigma_{z}=0.075$ 
are the uncertainties of the observed bias 
and redshift respectively (see Table I).

\noindent
{\bf (2)} Regarding the analysis of the growth-rate data we use
\be
\label{Likel}
\chi^{2}_{\rm gr}({\bf p}_{2})={\bf M} {\bf C}_{\rm cov}^{-1} {\bf M}^{T}
\ee
where ${\bf M}= \{f\sigma_{8,\rm
      obs}(z_{1})-f\sigma_{8}(z_{1},{\bf p}_{2}),...,f\sigma_{8,\rm
      obs}(z_{n})-f\sigma_{8}(z_{n},{\bf p}_{2})\}$ and 
${\bf C}^{-1}_{\rm cov}$ is the inverse covariance matrix 
\cite{Sarg}. The theoretical growth-rate is given by:
\be
f\sigma_{8}(z,{\bf p}_{2})=\sigma_{8}D(z)\Omega_{m}(z)^{\gamma(z)}\;.
\ee
The vectors ${\bf p}_{1}$ and ${\bf p}_{2}$ provide the
free parameters that enter in deriving the theoretical expectations.
The first vector includes the free parameters which are related to
the expansion and the environment of the parent dark matter halo in which
the LRGs DES galaxies live. 
Specifically, for constant $\gamma$ we have 
${\bf p}_{1}=\{{\bf p}_{2},M_{h}\}=
\{\Omega_{m0},h,\sigma_{8}\gamma,M_{h}\}$, while for the
case of evolving $\gamma$, the vector is defined 
as: ${\bf p}_{1}=\{\Omega_{m0},h,\sigma_{8},\gamma_0,\gamma_1,M_{h}\}$.
We remind the reader that the cosmological parameters 
$\{\Omega_{m0},h,\sigma_{8}\}=\{0.301,0.682,0.801\}$ 
are given in section II \cite{Abbot2017}.


Since we wish to perform a 
joint likelihood analysis of the two cosmological
probes and owing to the fact that likelihoods are defined
as ${\cal L}_{i}\propto {\rm exp}(-\chi^{2}_{i}/2)$, the 
overall likelihood function becomes 
\begin{equation}\label{eq:overalllikelihoo1}
{\cal L}({\bf p}_{1})=
{\cal L}_{\rm DES}
\times {\cal L}_{\rm gr}\;,
\end{equation}
which is equivalent to:
\begin{equation}\label{eq:overalllikelihoo}
\chi^{2}({\bf p}_{1})=
\chi^{2}_{\rm DES}+
\chi^{2}_{\rm gr}\;.
\end{equation}
Based on the above we will provide our results for each free parameter that
enters in the ${\bf p}_{1}$ vector. The uncertainty of each
fitted parameter will be estimated after marginalizing one
parameter over the other, providing as its uncertainty the
range for which $\Delta \chi^{2}(\le 1\sigma)$.

As a further quality measure over the fits, we have used 
the AIC \cite{Akaike1974} 
criterion, in a modified form that is appropriate for small data sets, \cite{Liddle:2007fy}. Considering Gaussian errors AIC is given by
\begin{eqnarray}
{\rm AIC} = -2 \ln {\cal L}_{\rm max}+2k+\frac{2k(k+1)}{N-k-1} \label{eq:AIC}\;,\\
\end{eqnarray}
where $N$ is the total number of data 
and $k$ is the number of fitted parameters (see also \cite{Liddle:2007fy}).
Of course, a smaller value of AIC implies a better model-data fit.
In order to test
the performance of the different bias models in fitting
the data we need to utilize the model
pair difference, namely
$\Delta {\rm AIC}={\rm AIC}_{\rm model}-{\rm AIC}_{\rm min}$.
From one hand, the restriction $\Delta {\rm AIC} \le 2$ 
indicates consistency between the two comparison models.
On the other hand, the inequalities
$4<\Delta {\rm AIC} <7$ indicate a positive evidence
against
the model with higher value of ${\rm AIC}_{\rm model}$ \cite{Ann2002, Burham2004},
while the condition 
$\Delta {\rm AIC} \ge 10$ suggests a strong such evidence.




\subsection{Observational constraints}
Below, we provide a qualitative discussion 
of our constraints, giving the reader the opportunity
to appreciate the new results of the current study.


\subsubsection{Constant growth index}
Here we focus on the 
$\Gamma_{1}$ parametrization, which means that the parameter 
space contains the following free parameters $(\gamma,M_{h})$.
The presentation of our constraints is provided in Table II
for the case of DES/Planck/JLA/BAO reference cosmology (see section II).
The Table includes the goodness of fit statistics 
($\chi_{\rm min}^2$, AIC), for the
specific bias models.
Also, in Figure 1 we present the 1$\sigma$, 2$\sigma$ and $3\sigma$
confidence contours in the $(\gamma,M_{h})$ plane. 

In particular, we find:

\begin{itemize}

\item For SMT model:
$\chi^2_{\rm min}=15.042$ (AIC=19.542),  $\gamma=0.640\pm 0.071$
and ${\rm log} (M_{h}/h^{-1}M_{\odot})= 13.000 \pm 0.072 $.

\item For JING model:
$\chi^2_{\rm min}=15.975$ (AIC=20.475),  $\gamma=0.650\pm 0.063$
and ${\rm log} (M_{h}/h^{-1}M_{\odot})= 12.910  \pm 0.062 $.

\item DMR model:
$\chi^2_{\rm min}=15.098$ (AIC=19.598),  $\gamma=0.640\pm 0.066$
and ${\rm log} (M_{h}/h^{-1}M_{\odot})= 12.730 \pm 0.074 $.

\item For BPR model:
$\chi^2_{\rm min}=17.048$ (AIC=21.548),  $\gamma=0.640\pm 0.075$
and ${\rm log} (M_{h}/h^{-1}M_{\odot})= 13.080  \pm 0.073 $.

\end{itemize}


We observe that the aforementioned bias models provide
very similar results (within $1\sigma$ errors) 
as far as the growth index 
is concerned. The corresponding 
best fit values show a small but non-zero deviation 
from the theoretically predicted value of GR
$\gamma_{\rm GR} \approx 6/11$
(see solid lines of Fig. 1), where the range of 
the confidence level is $\sim 1.3\sigma-1.7\sigma$. 
Such a small discrepancy between the predicted 
and observationally fitted value of $\gamma$ has also
been discussed by other authors. 
For example recently, \cite{Zhao} found $\gamma=0.656^{+0.042}_{-0.046}$, while 
\cite{Yi} obtained $\gamma=0.628^{+0.036}_{-0.039}$. 
Also, similar results can be found in previous papers \cite{Other} 
in which the tension can reach to $\sim 2.5 \sigma$.

Furthermore, we find that 
the best bias model is the SMT, however the inequality 
${\Delta}$AIC $\le 2$ indicates that 
the SMT bias model is statistically equivalent with rest of the 
models.
The second result is that the
differences of the bias models are absorbed in the fitted
value of the DM halo mass in which LRGs live, and which ranges from
$\sim 6.2\times 10^{12}h^{-1}M_{\odot}-1.2\times 10^{13}h^{-1}M_{\odot}$, for 
the different bias models and in
the case of DESY1COSMO bias.
As it can also be seen from Table II, our derived
mass of the
host DM halo mass is consistent with that of
Papageorgiou {\it et al}. \cite{Pap2018}, while 
Sawangwit {\it et al}. \cite{Sawangwit2011} and 
Pouri {\it et al}. \cite{Pouri2014} found 
$M_{h}\simeq (1.9-2) \times 10^{13}h^{-1}M_{\odot}$.

\begin{figure}[ht]
\includegraphics[width=0.5\textwidth]{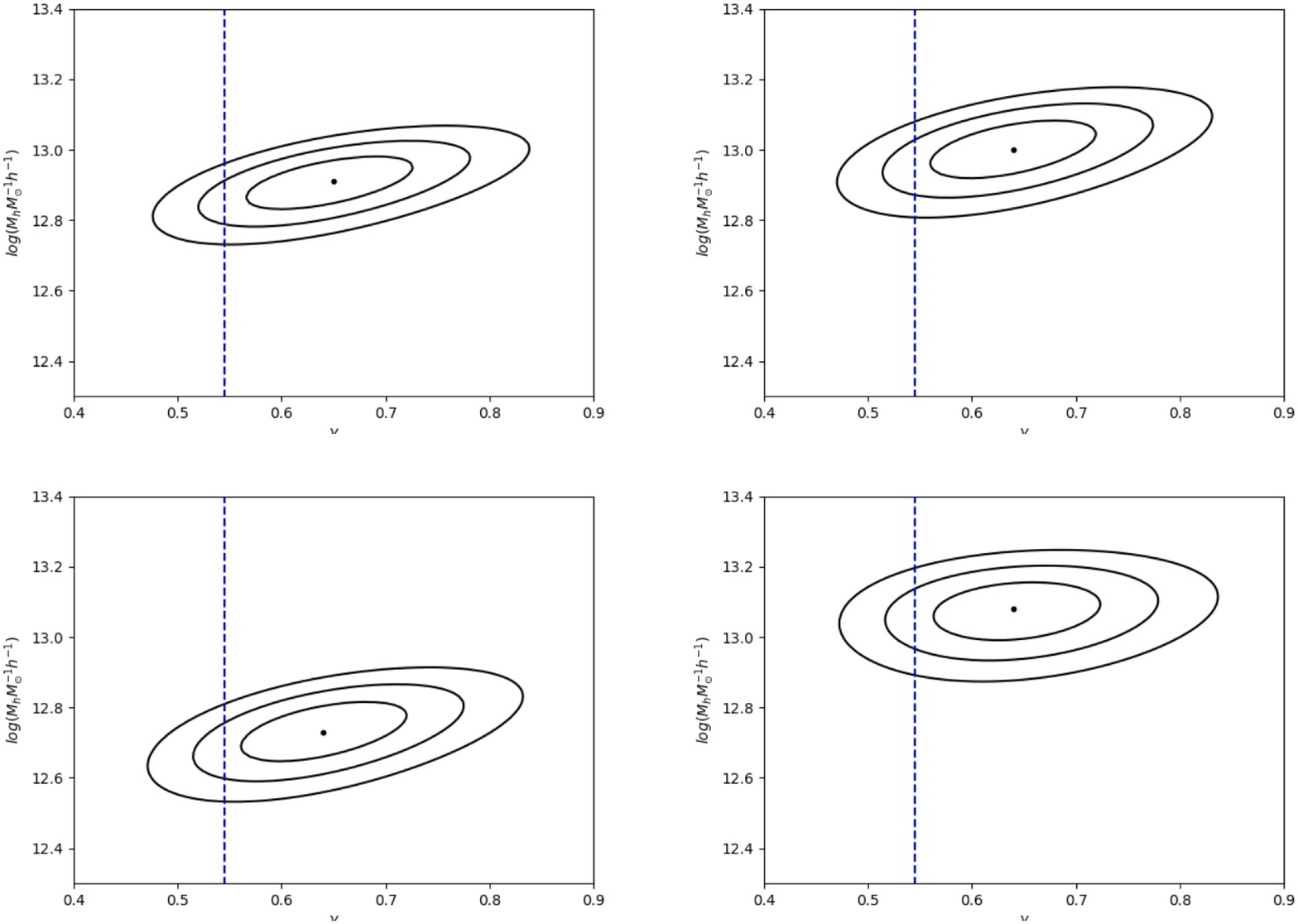}
\caption{The iso-likelihood contours for 
1$\sigma$ - 2$\sigma$-3$\sigma$ levels in the 
$(\gamma,M_h)$ parameter space for different bias models. 
\emph{Upper row:} From left to right, JING and SMT models. 
\emph{Lower row:} DMR and BPR models. 
For further details regarding the models, please see the 
relevant subsection. Notice that we use the 
DES/Planck/JLA/BAO $\Lambda$CDM cosmology \cite{Abbot2017}.
The best fit values are given in Table \ref{fig:DES_gamma_const}. 
The vertical dashed line corresponds to $\gamma_{\rm GR} \approx 
6/11$.  }\label{fig:DES_gamma_const}
\end{figure}

\begin{figure}[ht]
\includegraphics[width=0.5\textwidth]{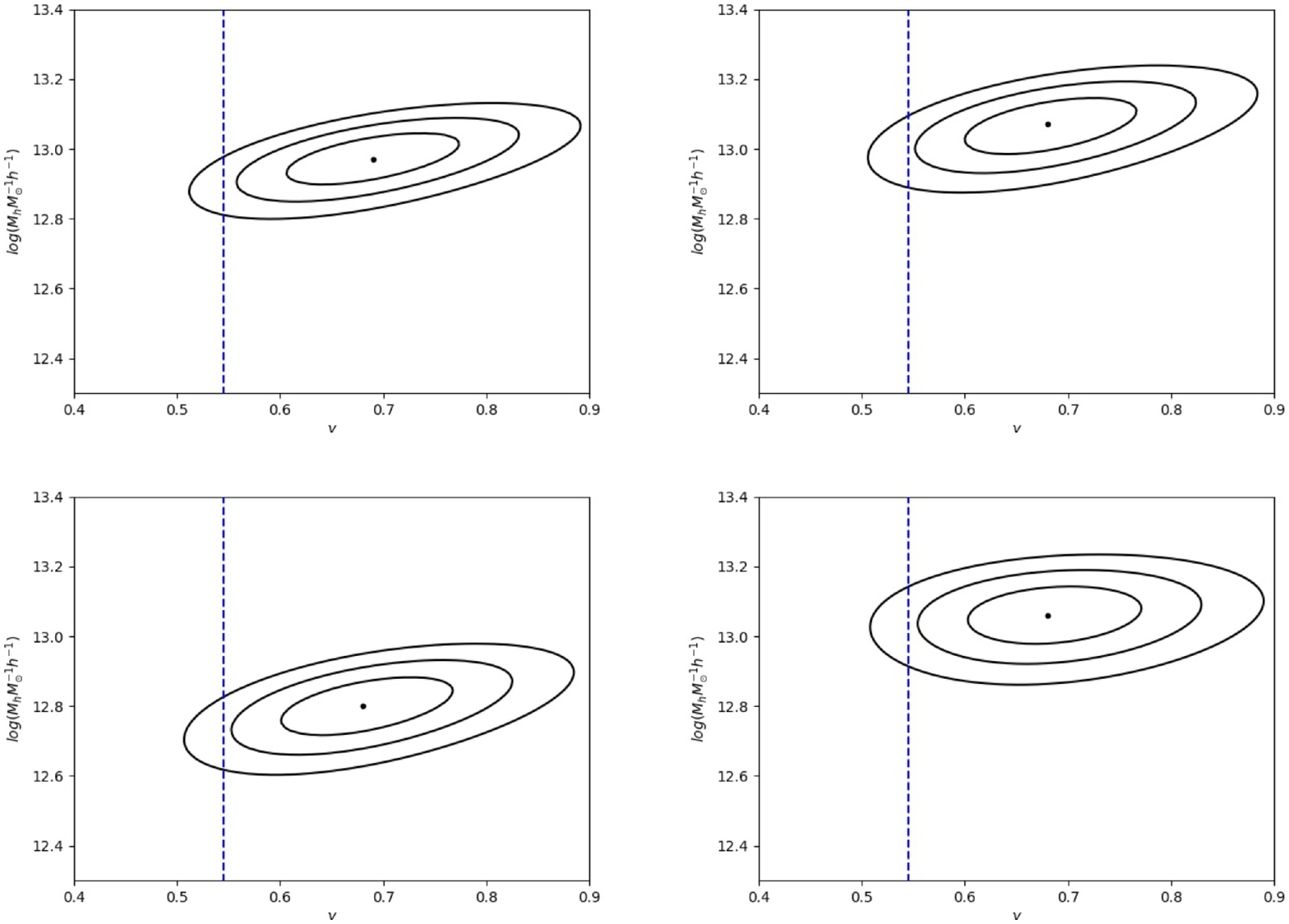}
\caption{The iso-likelihood contours for 
1$\sigma$ - 2$\sigma$-3$\sigma$ levels in the 
$(\gamma,M_h)$ parameter space for different bias models. 
\emph{Upper row:} From left to right, JING and SMT models. 
\emph{Lower row:} DMR and BPR models. 
For further details regarding the models, please see the 
relevant subsection. Notice that we use the 
Planck TT+TE+EE+low+lensing $\Lambda$CDM cosmology \cite{Aghanim:2018eyx}.
The best fit values are given in Table \ref{fig:DES_gamma_const}. 
The vertical dashed line corresponds to $\gamma_{\rm GR} \approx 
6/11$.  }\label{fig:DES_gamma_const}
\end{figure}


\begin{table*}
\caption[]{Observational constraints for the joint 
analysis of bias (see Table I) and growth rate data: The
  $1^{st}$ column shows the expansion model (see section II 
\cite{Abbot2017}), the
  $2^{nd}$ column indicates the bias models (see section III), 
the $3^{rd}$ column corresponds to $\gamma$ and the $4^{rth}$ provides the fitted DM halo mass. The remaining
  columns present the goodness-of-fit statistics $\chi^{2}_{\rm min}$,
  AIC and $\Delta$AIC$={\rm AIC}_{\rm,i}-{\rm
    AIC}_{\rm min}$. The index $i$ corresponds to the
  indicated bias model.}

\tabcolsep 4.0pt
\vspace{2mm}
\begin{tabular}{ccccccc} \hline \hline
$\Lambda$CDM Expansion Model & Bias Model & $\gamma$ & ${\rm log}(M/h^{-1}M_{\odot})$ &
  $\chi_{\rm min}^{2}$ &${\rm AIC}$&
  $\Delta$AIC \vspace{0.05cm}\\ \hline
DES/Planck/JLA/BAO (Abbott {\it et al}. \cite{Abbot2017}) & SMT & $0.640 \pm 0.071$ & $13.000 \pm 0.072$   &15.042& 19.542 & 0   
\vspace{0.01cm}\\
            &JING& $0.650 \pm 0.063$ & $12.910 \pm 0.062$   &15.975& 20.475 & 0.933   \vspace{0.01cm}\\
            &DMR& $0.640 \pm 0.066$ & $12.730 \pm 0.074$   &15.096& 19.598 & 0.056  \vspace{0.01cm}\\
            &BPR& $0.640 \pm 0.075$ & $13.080 \pm 0.073$   &17.048& 21.548 & 2.006  \vspace{0.15cm}\\
Planck TT+TE+EE+low+lensing (Aghanim {\it et al}. \cite{Aghanim:2018eyx}) & SMT & $0.680 \pm 0.076$ & $13.07 \pm 0.064$   &15.057& 19.557 & 0   
\vspace{0.01cm}\\
            &JING& $0.690 \pm 0.071$ & $12.97 \pm 0.058$   &15.952& 20.452 & 0.895   \vspace{0.01cm}\\
            &DMR& $0.680 \pm 0.075$ & $12.80 \pm 0.073$   &15.104& 19.604 & 0.047  \vspace{0.01cm}\\
            &BPR& $0.680 \pm 0.089$ & $13.06 \pm 0.080$   &16.947& 21.447 & 18.49  \vspace{0.01cm}\\
\hline\hline
\label{tab:growth1}
\end{tabular}
\end{table*}

In order to complete the present investigation 
we repeat the likelihood procedure in the case of  
Planck TT+TE+EE+low+lensing $\Lambda$CDM cosmology, hence
$\Omega_{m0}=1-\Omega_{\Lambda0}=0.3153$, $h=0.6736$, 
$\Omega_{b0}=0.0493$, $n=0.9649$, and $\sigma_{8}=0.811$
\cite{Aghanim:2018eyx}. Specifically, for the explored bias models 
we obtain (see also Table II):

\begin{itemize}

\item For SMT model:
$\chi^2_{\rm min}=15.057$ (AIC=19.557),  $\gamma=0.680\pm 0.076$
and ${\rm log} (M_{h}/h^{-1}M_{\odot})= 13.070  \pm 0.064$.

\item For JING model:
$\chi^2_{\rm min}=15.952$ (AIC=20.452),  $\gamma=0.690\pm 0.071$
and ${\rm log} (M_{h}/h^{-1}M_{\odot})= 12.970  \pm 0.058 $.

\item DMR model:
$\chi^2_{\rm min}=15.104$ (AIC=19.604),  $\gamma=0.680\pm 0.075$
and ${\rm log} (M_{h}/h^{-1}M_{\odot})= 12.800 \pm 0.073 $.

\item For BPR model:
$\chi^2_{\rm min}=16.947$ (AIC=21.447),  $\gamma=0.680\pm 0.089$
and ${\rm log} (M_{h}/h^{-1}M_{\odot})= 13.060  \pm 0.080 $.

\end{itemize}


Obviously, our statistical results remain quite robust (within $1\sigma$)
against the choice of the undelying expansion 
\cite{Aghanim:2018eyx}, \cite{Abbot2017}. Moreover, as it can be seen 
from Fig.2 the growth index of the 
Planck TT+TE+EE+low+lensing $\Lambda$CDM cosmology deviates 
with respect to that of GR ($\gamma_{\rm GR} \approx 6/11$) 
at $\sim 1.5-2\sigma$ levels.

\subsubsection{Constraints on $\gamma(z)$}
In this section we implement the overall likelihood procedure in the
$(\gamma_0,\gamma_1)$ parameter space. Based on the 
considerations discussed in the previous section
the statistical vector takes the form
${\bf p}_{1}=\{\Omega_{m0},h,\sigma_{8},\gamma_0,\gamma_1,M_{h}\}$.

In Fig.3 we plot the results of our statistical analysis in the
$(\gamma_{0},\gamma_{1})$ plane for the SMT bias model, since we
have verified that using the other bias models 
we get similar contours. 
The predicted $(\gamma_{0}^{(th)},\gamma_{1}^{(th)})$
$\Lambda$CDM values are indicated by the solid point, while
the star corresponds to our best fit values.
In brief for the DES/Planck/JLA/BAO $\Lambda$ cosmology
we find
$\gamma_0=0.630 \pm 0.072$, $\gamma_1=0.040 \pm 0.403$ with 
$\chi^2_{\rm min}=15.033$ (AIC=19.533), while 
in the case of Planck TT+TE+EE+low+lensing $\Lambda$CDM cosmology
we get 
$\gamma_0=0.670 \pm 0.073$, $\gamma_1=0.100 \pm 0.422$ with 
$\chi^2_{\rm min}=14.988$ (AIC=19.488).
Notice that for the sake of simplicity 
we have marginalized the likelihood analysis over
the LRG dark matter halo, namely 
${\rm log} (M_{h}/h^{-1}M_{\odot})= 13.00$ and 13.07 respectively 
(see SMT model in Table II).

We conclude that the joint statistical analysis put tight 
constraints $\gamma_{0}$, however for $\gamma_{1}$ the corresponding
error bars remain quite large. Also the range of deviation from GR is   
$1.7-2.9\sigma$.
We argue that with the next generation of data (mainly from {\it Euclid})
we will be able 
to test whether the growth index of matter fluctuations depends on time.

\begin{figure}[ht]
\includegraphics[width=0.5\textwidth]{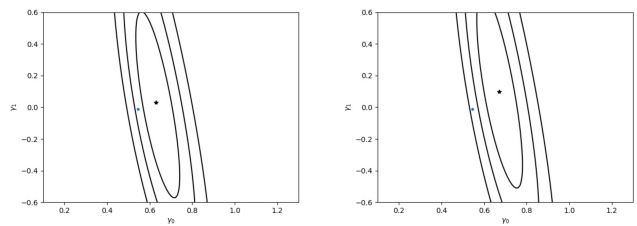}
\caption{Iso - likelihood contours for 
$\Delta \chi^{2}=-2{\rm ln}{\cal L}/{\cal L}_{\rm max}$ equal to 2.30, 
6.18 and 11.83, corresponding
to 1$\sigma$, 2$\sigma$ and $3\sigma$ confidence levels in the
$(\gamma_{0},\gamma_{1})$ plane in the case of 
$\Gamma_{1}$ parametrization. The bias model is that of SMT, while 
the star corresponds to the best-fit point and the dot to the theoretical $\Lambda$CDM point
$(\gamma_{0},\gamma_{1})=(0.556,-0.011)$.
In the right panel we 
present the contours that obtained using the Planck TT+TE+EE+low+lensing $\Lambda$CDM cosmology \cite{Aghanim:2018eyx}, while in the left the contours obtained using DES/Planck/JLA/BAO $\Lambda$CDM cosmology.
}\label{fig:SMT_g0g1_res}
\end{figure}

\section{Conclusions}
In the context of the concordance $\Lambda$CDM model, 
testing the validity of general relativity (GR) on extragalactic scales 
is considered one of the most important tasks in cosmological studies, 
hence it is crucial to minimize the amount of priors needed
to successfully complete such an effort. 
One such prior is the growth index ($\gamma$) of matter perturbations. 
It is well known that a 
necessary step toward testing GR is to measure $\gamma$ at the 
$\sim 1\%$ accuracy level. Obviously, in order to 
control the systematic effects that 
possibly affect individual methods and tracers of the growth
of matter perturbations we need to 
have independent estimations of $\gamma$. 

Despite the fact that the $\Lambda$CDM+GR model is found to be 
in a very good agreement with the majority of cosmological 
data~\citep{Aghanim:2018eyx}, nonetheless the model seems 
to be currently in tension with some recent measurements~\citep{s8}, 
related with the Hubble constant $H_0$ and 
the present value of the mass variance at 8$h^{-1}$Mpc $\sigma_8$. 
Whether these tensions are the result of yet 
unknown systematic errors or hint some underlying
new Physics is still unclear. 
In the light of the latter results, an intense debate 
is taking place in the literature
and the aim of the present article is to contribute to this debate.

In this article we used the biasing properties of 
the Luminous Red Galaxies, recently released by the group of 
Dark Energy Survey (DES), together with growth rate data 
in order to constrain the growth index of matter perturbations.
Specifically, in the framework of concordance $\Lambda$ cosmology, 
we study the ability of four bias models to 
fit the DES bias data. Then we combined bias 
in a joint analysis with the growth rate of matter fluctuations
to place constraints on the parameters. 

Considering a
constant growth index we placed constraints, up to
$\sim 10\%$ accuracy, on the growth index. 
Specifically, using the priors of the Dark Energy Survey
we found that the constraints 
remain mostly unaffected by using different forms of bias.
In particular, we obtained $0.640 \pm 0.071$, $0.650 \pm 0.063$, $0.640 \pm 0.066$ and   
$\gamma=0.640\pm 0.075$ for SMT \cite{She2001}, JING \cite{Jing1998}, DMR \cite{deSim2011}  
and BPR \cite{Basilakosetal2012} bias models. Also 
utilizing the Planck priors we got 
$\gamma=0.680\pm 0.076$, $\gamma=0.690\pm 0.071$, $\gamma=0.680\pm 0.075$
and $\gamma=0.680\pm 0.089$ for the aforementioned bias factors.
Obviously, we found a small but non-zero deviation from 
GR ($\gamma_{\rm GR}\approx 6/11$), where the confidence level 
lies in the interval $\sim 1.3\sigma-2\sigma$.
Such a small discrepancy between the predicted 
and observationally fitted value of $\gamma$ has also
been reported in several studies 
\cite{Zhao}, \cite{Yi} and \cite{Other}. 
Moreover, the intrinsic differences of the bias
models are absorbed in the fitted value of the
dark-matter halo mass in which LRGs survive, and which
belongs in the range $\sim 6.2\times 10^{12}h^{-1}M_{\odot}-1.2\times 10^{13}h^{-1}M_{\odot}$.

Under the assumption that the growth index varies with time, namely 
$\gamma(z)=\gamma_{0}+\gamma_{1}z/(1+z)$, 
we showed that the $(\gamma_{0},\gamma_{1})$ parameter solution space 
accommodates the GR $(\gamma_{0},\gamma_{1})$ values at $\sim 1.7\sigma$ ($\sim 2.9 \sigma$) level utilizing the DES/Planck/JLA/BAO (Planck) priors.
Similar to previous studies, we placed tight 
constraints on $\gamma_{0}$, however the corresponding uncertainties 
of $\gamma_{1}$ remain large.  
The next generation (mainly from {\it Euclid}) of dynamical data are
expected to improve the constraints on $\gamma_{1}$, hence 
the validity of general relativity on extragalactic scales 
will be effectively checked.

\section*{Acknowledgments}  
Spyros Basilakos acknowledges support from the Cyprus Research Promotion Foundation 
in the context of the program “GRATOS: Graph Theoretical Tools for Sciences” (ref. number EXCELLENCE/216/0207) led by European University Cyprus.


\begin{thebibliography}{plain}


\bibitem{Hicken2009}
M. Hicken {\it et al}., Astrophys. J., {\bf 700}, 1097 (2009)

\bibitem{Komatsu2011}
E. Komatsu {\it et al}., Astrophys. J. Supp., {\bf 192}, 18 (2011)

\bibitem{Blake2011}
C. Blake {\it et al}., Mon. Not. Roy. Soc., {\bf 418}, 1707 (2011)

\bibitem{Hinshaw2013}
G. Hinshaw  {\it et al}., Astrophys. J. Supp., {\bf 208}, 19 (2013)

\bibitem{Farooq2013}
O. Farooq, D. Mania and B. Ratra, Astrophys. J., {\bf 764}, 138 (2013).

\bibitem{Ade2013}
Planck Collab. 2015, P.A.R. Ade {\it et al}., Astron. Astrophys. {\bf 594} (2016) A13. 


\bibitem{Aghanim:2018eyx}
  N.~Aghanim {\it et al.} [Planck Collaboration],
  arXiv:1807.06209 [astro-ph.CO].

\bibitem{Copeland2006}
E. J. Copeland, M. Sami and S. Tsujikawa, Int. J. of Mod. Phys. D., {\bf 15}, 1753 (2006)

\bibitem{Caldwell2009}
R. R. Caldwell and M. Kamionkowski, Ann. Rev. Nucl. Part. Sci., {\bf 59}, 397 (2009)

\bibitem{Amendola2010}
L. Amendola and S. Tsujikawa, Dark Energy: Theory and Observations, Cambridge University Press, Cambridge UK (2010)




\bibitem{Linder2004}
E. V. Linder, Phys. Rev. D., {\bf 70}, 023511 (2004)  

\bibitem{Linder2007}
E. V. Linder and R. N. Cahn, Astrop. Phys., {\bf 28}, 481 (2007)  

\bibitem{Mar2014}
H. Steigerwald, J. Bel and C. Marinoni, JCAP, {\bf 5}, 42, 2014 


\bibitem{Bas2016}
S. Basilakos \& S. Nesseris, Phys. Rev. D., {\bf 84}, 123525 2016;
Phys. Rev. D. {\bf 96}, 063517 2017

\bibitem{Peebles1993}
P. J. E. Peebles, ``Principles of Physical
Cosmology'', Princeton University Press, Princeton New Jersey (1993)

\bibitem{Wang1998}
L. Wang and P. J. Steinhardt, Astrophys. J., {\bf 508}, 483 (1998)


\bibitem{Ness2015}
S. Nesseris, D. Sapone, and J. Garcia-Bellido, Phys. Rev. D.,
{\bf 91}, 023004 (2015)


\bibitem{Silveira1994}
V. Silveira and I. Waga, Phys. Rev. D., {\bf 50}, 4890 (1994)

\bibitem{Nesseris2008}
S. Nesseris and L. Perivolaropoulos, Phys. Rev. D.,
{\bf 77}, 023504 (2008)

\bibitem{Basilakos2012}
S. Basilakos, Intern. Journal of Modern Physics D, {\bf 21}, 1250064 (2012);
S. Basilakos and A. Pouri, Mon. Not. Roy. Soc., {\bf 423}, 3761 (2012)


\bibitem{Wei2008}
H. Wei, Phys. Lett. B., {\bf 664}, 1 (2008)

\bibitem{Gong2008}
Y. Gong, Phys. Rev. D., {\bf 78}, 123010 (2008)

\bibitem{Fu2009}
X. Fu, P. Wu and H. Yu, Phys. Lett. B, {\bf 677}, 12 (2009)

\bibitem{Gannouji2009}
R. Gannouji, B. Moraes and D. Polarski, JCAP, {\bf 2}, 34 (2009)

\bibitem{Tsujikawa2009}
S. Tsujikawa, R. Gannouji, B. Moraes and D. Polarski, Phys. Rev. D., {\bf 80}, 084044 (2009)

\bibitem{BasFT}
S. Basilakos, Phys. Rev. D., {\bf 93}, 083007 (2016)

\bibitem{Basilakos2013}
S. Basilakos and P. Stavrinos, Phys. Rev. D., {\bf 87}, 043506 (2013);
G. Papagiannopoulos, S. Basilakos, A. Paliathanasis, S. 
Savvidou and P. C. Stavrinos, Class. Quant. Grav., {\bf 34}, 225008 (2017)

\bibitem{Basola2015}
 S.  Basilakos  and  J.  Sola,  Phys.  Rev.  D.,
{\bf 92},  123501 (2015) 

\bibitem{Mehra2015}
 A. Mehrabi, S. Basilakos and F. Pace, Mon. Not. R. Soc, {\bf 452}, 2930 (2015); A. Mehrabi,  S. Basilakos,  M. Malekjani and Z. Davari, Phys. Rev. D.
{\bf 92}, 123513 (2015)

\bibitem{Matsubara2004}
T. Matsubara, Astrophys. J., {\bf 615}, 573 (2004)

\bibitem{Basilakos2005}
S. Basilakos and M. Plionis, Mon. Not. Roy. Soc., {\bf 360}, L35 (2005)

\bibitem{Basilakos2006}
S. Basilakos and M. Plionis, Astrophys. J., {\bf 650}, L1 (2006)

\bibitem{Krumpe2013}
M. Krumpe, T. Miyaji and A. L. Coil, arXiv:1308.5976 (2013)


\bibitem{Basilakosetal2012}
S. Basilakos, M. Plionis and A. Pouri, Phys. Rev. D., {\bf 83}, 
123525 (2011); 
S. Basilakos, J. B. Dent, S. Dutta, L. Perivolaropoulos and M. Plionis, Phys. Rev. D., {\bf 85}, 123501 (2012);
S. Basilakos and M. Plionis, Astrophys. J., {\bf 550}, 522 (2001)


\bibitem{Bean2013}
R. Bean {\it et al}., arXiv:1309.5385 (2013)


\bibitem{DES2017} 
M. Elvin-Poole, {\it et al}., Phys. Rev. D {\bf 98}, 042006 (2018)
(DES Y1COSMO)

\bibitem{Krause2017} 
Krause, E., {\it et al}. (DES Collaboration), submitted to Phys.
Rev. D (2017), arXiv:1706.09359 [astro-ph.CO]

\bibitem{Abbot2017}
T. M. C. Abbott {\it et al}., Phys. Rev. D {\bf 98}, 043526 (2018)


\bibitem{Sarg}
B. Sargedo, S. Nesseris and D. Sapone, Phys. Rev. D., {\bf 98}, 083543 (2018)


\bibitem{Song09}
Y-S. Song and W.J. Percival, JCAP, {\bf 10}, 4, (2009)



\bibitem{She2001},  
R. K. Sheth and G. Tormen, G. 1999, Mon. Not. R. Astron. Soc., 
{\bf 323} 1 (2001)

\bibitem{Jing1998} Y. P. Jing, ApJ, {\bf 503}, L9 (1998)


\bibitem{deSim2011} A. de Simone, M. Maggiore, 
A. Riotto, Mon. Not. R. Astron. Soc., {\bf 412}, 2587 (2011)

\bibitem{Wein} 
N. N. Weinberg and M. Kamionkowski, Mon. Not. R. Astron. Soc.,
{\bf 341}, 251 (2003)


\bibitem{EisensteinHu} 
D. J. Eisenstein and W. Hu, ApJ, {\bf 496}, 605 (1998)

\bibitem{Sugiyama1995}
N. Sugiyama, ApJ, {\bf 100}, 281 (1995)



\bibitem{Lue2004}
A. Lue, R. Scoccimarro and G. D. Starkman, Phys. Rev. D., {\bf 69}, 124015 (2004)

\bibitem{Stabenau2006}
H. F. Stabenau and B. Jain, Phys. Rev. D., {\bf 74}, 084007 (2006)

\bibitem{Uzan2007}
P. J. Uzan, Gen. Rel. Grav., {\bf 39}, 307 (2007)   

\bibitem{Tsujikawa2008}
S. Tsujikawa, K. Uddin and R. Tavakol, Phys. Rev. D., {\bf 77}, 043007 (2008)


\bibitem{Polarski2008}
D. Polarski and R. Gannouji, Phys. Lett. B., {\bf 660}, 439 (2008)

\bibitem{Bal08}
G. Ballesteros and A. Riotto, Phys. Lett. B. {\bf 668}, 171 (2008)

\bibitem{Belloso2009}
 A. B. Belloso, J. Garcia-Bellido and D. Sapone, JCAP, {\bf 10}, 10 (2011)

\bibitem{Akaike1974}
H. Akaike,
{\it{A new look at the statistical model identification}},
\href{https://ieeexplore.ieee.org/
document/1100705/}{\emph{IEEE Transactions
on Automatic Control},
\textbf{19}, (1974)  716}.



\bibitem{Liddle:2007fy}
  A.~R.~Liddle,
  \emph{Information criteria for astrophysical model selection},
\href{https://onlinelibrary.wiley.
com/doi/full/10.1111/j.1745-3933.2007.00306.x}{\emph{Mon.\ Not.\ Roy.\ Astron.\ Soc.\
}{\bfseries
377}, (2007) L74}, [\href{https://arxiv.org/abs/astro-ph/0701113}{0701113}].

\bibitem{Ann2002}
 K. P. Burnham, D. R. Anderson,
{\it{Model selection and multimodel inference:
a practical information-theoretic approach}},
\href{https://www.springer.com/gp/book/9780387953649}{
2nd edn. Springer,
New York (2002)}

\bibitem{Burham2004}
 K. P. Burnham, D. R. Anderson,
\emph{Multimodel inference: Understanding AIC and BIC in Model Selection},
\href{journals.sagepub.
com/doi/abs/10.1177/0049124104268644}{\emph{Sociol. Meth. \& Res.,} {\bfseries 33}, (2004)
 261}

\bibitem{Zhao}
Ming-Ming  Zhao, Jing-Fei  Zhang and  Xin  Zhang, 
Physics Lett. B {\bf 779} 473 (2018) 

\bibitem{Yi}
Zhao-Yu Yin and Hao Wei, (2019), [arXiv:1902.00289]

\bibitem{Other}
 L. Xu, Phys. Rev. D {\bf 88}, 084032 (2013);
H. Gil-Marn, W. J. Percival, L. Verde, J. R. Brownstein,
C. H. Chuang, F. S. Kitaura, S. A. Rodrguez-Torres and
M. D. Olmstead, Mon. Not. Roy. Astron. Soc.
{\bf 465}, 1757 (2017); L.  Samushia
{\it et al}.,  Mon.  Not.  Roy.  Astron.  Soc.
{\bf 429},
1514 (2013); F. Beutler
{\it et al}.
[BOSS Collaboration], Mon. Not. Roy. Astron.  Soc.
{\bf 443} 1065  (2014); S. Basilakos,  Mon. Not. Roy. Astron.  Soc.,
{\bf 449}, 2151 (2015)


\bibitem{Pap2018}
	A. Papageorgiou, S. Basilakos and M. Plionis, Mon. Not. R. Astron. Soc., {\bf 476}, 2621 (2018)


\bibitem{Sawangwit2011}  
U. Sawangwit, {\it et al}. Mon. Not. R. Astron. Soc. {\bf 416} 3033

\bibitem{Pouri2014}
A. Pouri, S. Basilakos and M. Plionis, JCAP, 
J. Cosmol. \& Astrop. Phys., {\bf 08}, 042, (2014)  

\bibitem{s8} For recent reviews see:  L.~Verde, T.~Treu and A.~G.~Riess,
  Nature Astronomy 2019
  doi:10.1038/s41550-019-0902-0
  [arXiv:1907.10625 [astro-ph.CO]], and references therein;
 J.~Sol\`a Peracaula,
  Int.\ J.\ Mod.\ Phys.\ A {\bf 33}, no. 31, 1844009 (2018).
  doi:10.1142/S0217751X18440098, and references therein. 



\end{thebibliography}
\end{document}